\documentclass[runningheads]{rnl}

\usepackage[rightcaption]{sidecap}

\usepackage{graphicx, subfigure}  

\usepackage{amsmath,amsfonts,amssymb}   

\begin{document}

\title*{Short-distance propagation of nonlinear optical pulses}


\titrecourt{Propagation of nonlinear optical pulses}


\author{Mathieu Isoard\inst{1}, A. M. Kamchatnov\inst{2,3}\and
N. Pavloff\inst{1}}

\index{Isoard Mathieu}              
\index{Kamchatnov Anatoly}          
\index{Pavloff Nicolas}
\auteurcourt{M. Isoard {\it et al.}}

\adresse{LPTMS, UMR 8626, CNRS, Univ. Paris-Sud, Universit\'e
  Paris-Saclay, 91405 Orsay, France \and Institute of Spectroscopy,
  Russian Academy of Sciences, Troitsk, Moscow, 108840, Russia \and
  Moscow Institute of Physics and Technology, Institutsky lane 9,
  Dolgoprudny, Moscow region, 141701, Russia}

\email{mathieu.isoard@u-psud.fr}

\maketitle

\begin{resume}
  Nous \'etudions la propagation transverse d'un pulse lumineux
  quasi-unidimensionnel dans un milieu optique non-lin\'eaire, en
  pr\'esence d'un fond d'intensit\'e lumineuse constante.  Dans un
  premier temps, le signal initial se divise en deux parties qui se
  propagent dans des directions oppos\'ees. Ce ph\'enom\`ene peut
  \^etre d\'ecrit th\'eoriquement \`a l'aide d'une approche non
  dispersive en utilisant une modification de la m\'ethode de Riemann
  propos\'ee par Ludford. Les r\'esultats sont en excellent accord avec
  les simulations num\'eriques.
\end{resume}

\begin{resumanglais}
  We theoretically describe the quasi one-dimensional transverse
  spreading of a light pulse propagating in a defocusing nonlinear optical
  material in the presence of a uniform background light
  intensity. For short propagation distances the pulse can be
  described within a nondispersive approximation by means of Riemann's
  approach. The theoretical results are in excellent agreement with
  numerical simulations.
\end{resumanglais}

\section{Introduction}

It has long been realized that light propagating in a nonlinear medium
was amenable to a hydrodynamic treatment
\cite{isoard_Tal65,isoard_ASK67}. In the present work we use such an
approach to study a model configuration which has been realized
experimentally in a one-dimensional situation in the defocusing regime
in Ref.~\cite{isoard_wkf-07}: the nonlinear spreading of a region of
increased light intensity in the presence of a uniform constant
background. In the absence of background, and for a smooth initial
intensity pattern, the spreading is mainly driven by the nonlinear
defocusing and can be treated analytically in some simple cases
\cite{isoard_Tal65}. The situation is more interesting in the presence
of a constant background: the pulse splits in two parts, each
eventually experiencing nonlinear wave breaking, leading to the
formation of dispersive shock waves at both extremities of the
split pulse. In the present work we concentrate on the pre-shock
period and demonstrate that it can be very accurately described within
a non-dispersive nonlinear approximation.

The paper is organized as follows: In Sec.~\ref{isoard_model} we
present the model and the set-up we aim at studying. The spreading and
the splitting stage of evolution is accounted for in
Sec.~\ref{isoard_DSE} within a dispersionless approximation which
holds when the pulse region initially presents no large intensity
gradient. The problem is first mapped onto an Euler-Poisson equation
in Sec.~\ref{isoard_RV}. This equation is solved in
Sec.~\ref{isoard_Sol_EP} by using Riemann-Ludford method. In
Sec.~\ref{isoard_Results} the theoretical results are compared with
numerical simulations. Our conclusions are presented in
Sec.~\ref{isoard_conclusion}.

\section{The model} \label{isoard_model}

In the paraxial approximation, the stationary propagation of the
complex amplitude $A(\vec{r})$ of the electric field of a
monochromatic beam is described by the equation (see, e.g.,
Ref.~\cite{isoard_LL8})
\begin{equation*}
{\rm i}\partial_z A = -\frac{1}{2 n_0 k_0} \vec{\nabla}^2_{\!\perp} A
-k_0 \delta n\, A\; .
\end{equation*}
In this equation, $n_0$ is the linear refractive index,
$k_0=2\pi/\lambda_0$ is the carrier wave vector, $z$ is the longitudinal
coordinate
along the beam, $\vec{\nabla}^2_\perp$ the transverse Laplacian and
$\delta n$ is a nonlinear contribution to the index. In a non
absorbing defocusing Kerr nonlinear medium one can write
$\delta n=-n_2 |A|^2$, with $n_2>0$.

We consider a system with a uniform background light intensity,
denoted as $I_0$, on top of which an initial pulse is added at the
entrance of the nonlinear cell. To study the propagation of this pulse
along the beam (direction $z$), we introduce the following
characteristic quantities: the nonlinear length
$z_{\rm\scriptscriptstyle NL}=(k_0 n_2 I_{0})^{-1}$ and the transverse
healing length
$\beta_\perp=(z_{\rm\scriptscriptstyle NL}/n_0 k_0)^{1/2}$. Since the
transverse profile depends on a single Cartesian coordinate, we write
$\vec{\nabla}^2_{\!\perp} = \beta_\perp^{-2} \partial_x^2$ where $x$
is the dimensionless transverse coordinate, and also define an
effective ``time'' $t=z/z_{\rm\scriptscriptstyle NL}$. In this
framework, the quantity $\psi(x,t) = A(x,t)/\sqrt{I_0}$ is solution of
the dimensionless nonlinear Schr\"odinger (NLS) equation
\begin{equation}\label{isoard_eq:nls}
{\rm i}\,\psi_t=-\tfrac12 \psi_{xx}+|\psi|^2\psi\; .
\end{equation}
The initial $\psi(x,t=0)$ is real (i.e., no
transverse velocity or, in optical context, no focusing of the light
beam at the input plane), with a dimensionless intensity
$\rho(x,t)=|\psi|^2$ which departs from the constant background value
(which we denote as $\rho_0$) only in the region near the origin where
it forms a bump. To be specific, we consider the typical case where
\begin{equation}\label{isoard_rho_init}
\rho(x,0)= \rho_0 +\rho_1 \exp{(-2\,x^2/x_0^2)} \; ,\quad\mbox{and}\quad
u(x,0)=0\; .
\end{equation}
The maximal density of the initial profile is
$\rho(0,0)=\rho_0+\rho_1\equiv\rho_{m}$.

\section{The dispersionless stage of evolution} \label{isoard_DSE}

The initial pulse splits into two signals propagating in 
opposite directions of $x$ axis. The aim of this section is to theoretically
describe this splitting within a dispersionless approximation.

\subsection{Riemann variables and Euler-Poisson
  equation}\label{isoard_RV}
By means of the Madelung substitution
$\psi(x,t)=\sqrt{\rho(x,t)}\exp\left({\rm i} \int^x u(x',t)\,
  dx'\right)$,
the NLS equation \eqref{isoard_eq:nls} can be cast into a
hydrodynamic-like form for the density $\rho(x,t)$ and the flow
velocity $u(x,t)$:
\begin{equation}\label{isoard_3-3}
    \rho_t+(\rho u)_{x}=0\; ,\quad
    u_t+u u_{x}+\rho_{x}+\left(\frac{\rho_{x}^2}{8\rho^2}
    -\frac{\rho_{xx}}{4\rho}\right)_{x}=0 \; .
\end{equation}
These equations are to be solved with the initial conditions
\eqref{isoard_rho_init}.  The last term of the left
hand-side of the second of Eqs.~(\ref{isoard_3-3}) accounts for the
dispersive character of the fluid of light. In the first stage of
spreading of the bump, if the density gradients of the initial density
are weak (i.e., if $x_0\gg \mathrm{min}\{\rho_0^{-1/2},\rho_1^{-1/2}\}$), 
the effects of dispersion can be
neglected, and the system (\ref{isoard_3-3}) simplifies to
\begin{equation}\label{isoard_3-3a}
\rho_t+(\rho u)_x=0 \; , \quad u_t +u u_x + \rho_x=0 \; .
\end{equation}
The above equations can be written in a more symmetric form by
introducing the Riemann invariants
\begin{equation}\label{isoard_3-3b}
\lambda^{\pm}(x,t)=\tfrac12 u(x,t)\pm\sqrt{\rho(x,t)} \; ,
\end{equation}
which evolve according to the system [equi\-va\-lent to
(\ref{isoard_3-3a})]:
\begin{equation}\label{isoard_3-3c1}
\partial_t\lambda^\pm +v_\pm(\lambda^-,\lambda^+)\,
\partial_x\lambda^\pm=0 \; ,
\quad\mbox{with}\quad
  v_\pm(\lambda^-,\lambda^+)
  =\tfrac{1}{2}(3\lambda^\pm+\lambda^\mp)=u\pm\sqrt{\rho} \; .
\end{equation}
The Riemann velocities $v_\pm$ in \eqref{isoard_3-3c1} have a simple physical
interpretation for a smooth velocity and density distribution: $v_+$
($v_-$) corresponds to a signal which propagates downstream (upstream)
at the local velocity of sound $c=\sqrt{\rho}$ and which is dragged by
the background flow $u$.

The system \eqref{isoard_3-3c1} can be linearized by means of the hodograph
transform (see, e.g., Ref.~\cite{isoard_kamch-2000}) which consists in
considering $x$ and $t$ as functions of $\lambda^+$ and
$\lambda^-$. One readily obtains
\begin{equation}\label{isoard_eq:hodo1}
\partial_\pm x -v_\mp \partial_\pm
t=0,
\end{equation}
where $\partial_{\pm} \equiv \partial/ \partial \lambda^{\pm}$.
One introduces two auxiliary (yet unknown) functions
$W_{\pm}(\lambda^-,\lambda^+)$ such that
\begin{equation}
x - v_\pm(\lambda^-,\lambda^+)\,t = W_\pm(\lambda^-,\lambda^+).
\label{isoard_hodograph}
\end{equation}
Inserting the above expressions in \eqref{isoard_eq:hodo1} shows that the
$W_{\pm}$'s are solution of
Tsarev equations \cite{isoard_Tsa91}
\begin{equation}
\frac{\partial_-  W_+ }{ W_+ -  W_-} =
\frac{\partial_- v_+}{v_+-v_-}, \quad \text{and} \quad
\frac{\partial_+  W_- }{ W_+ -  W_-} = \frac{\partial_+ v_-}{v_+-v_-}.
\label{isoard_Tsarev}
\end{equation}
From Eqs.~\eqref{isoard_3-3c1} and \eqref{isoard_Tsarev} one can
verify that $\partial_- W_+ = \partial_+ W_-$, which shows that $W_+$
and $W_-$ can be sought in the form
\begin{equation}\label{isoard_eq:pot}
  W_\pm = \partial_\pm \chi,
\end{equation}
where $\chi(\lambda^-,\lambda^+)$ plays the role of a
potential. Substituting expressions \eqref{isoard_eq:pot} in one of
the Tsarev equations shows that $\chi$ is a solution of the following
Euler-Poisson equation
\begin{equation}\label{isoard_eq:EP}
\frac{\partial^2 \chi}{\partial \lambda^+ \partial\lambda^-}
- \frac{1}{2\,(\lambda^+ - \lambda^-)}
\left( \frac{\partial \chi}{\partial \lambda^+} -
\frac{\partial \chi}{\partial \lambda^-} \right) = 0 \; .
\end{equation}

\subsection{Solution of the Euler-Poisson
  equation} \label{isoard_Sol_EP}

We use Riemann's method (see, e.g., Ref.~\cite{isoard_Som64}) to solve the
Euler-Poisson equation \eqref{isoard_eq:EP} in the ($\lambda^+$,
$\lambda^-$)--plane which we denote below as the ``characteristic
plane''. We follow here the procedure exposed in
Ref.~\cite{isoard_Lud52} which applies to
non-monotonous initial distributions, such as the one corresponding to
Eq.~\eqref{isoard_rho_init}.

\begin{figure}[h!]
\centering
\includegraphics[scale=0.58]{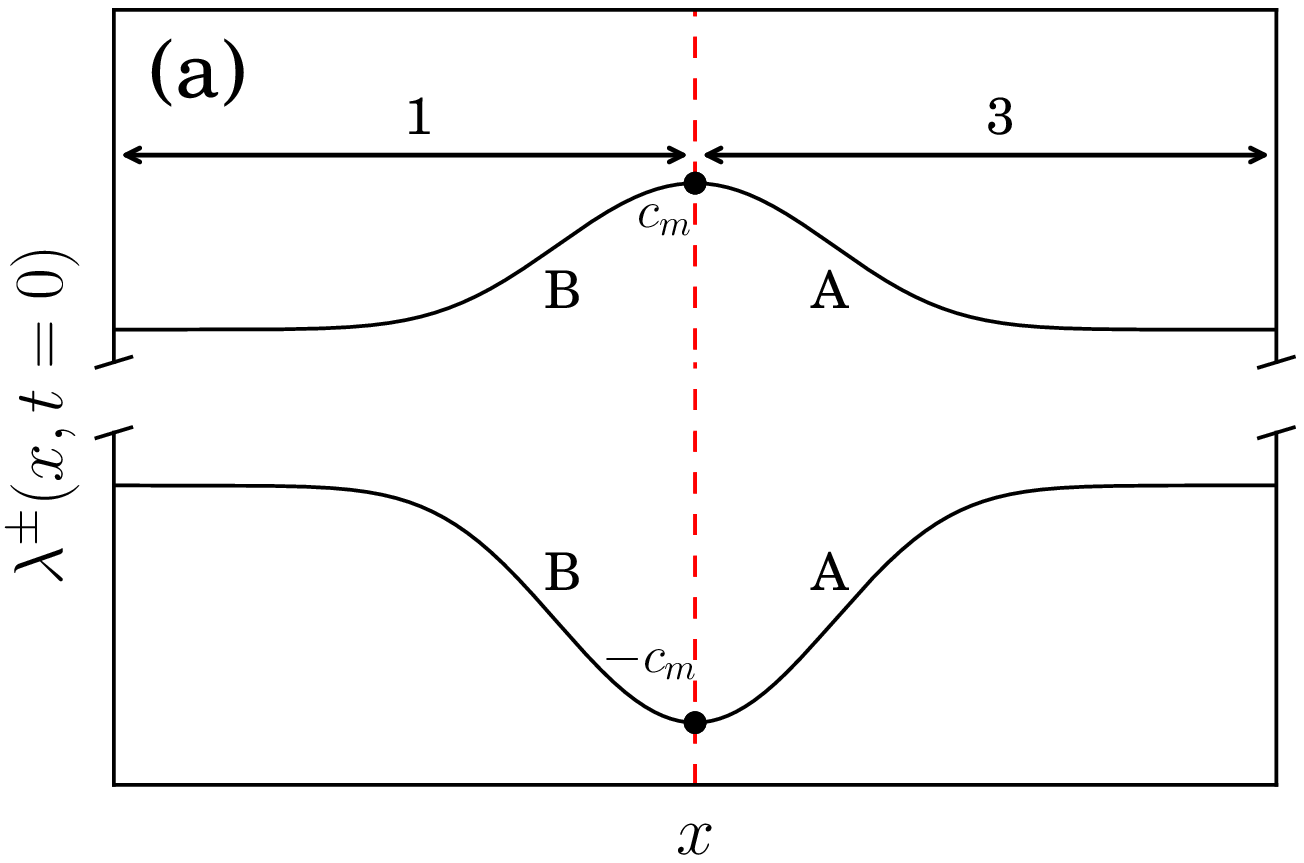}
\includegraphics[scale=0.58]{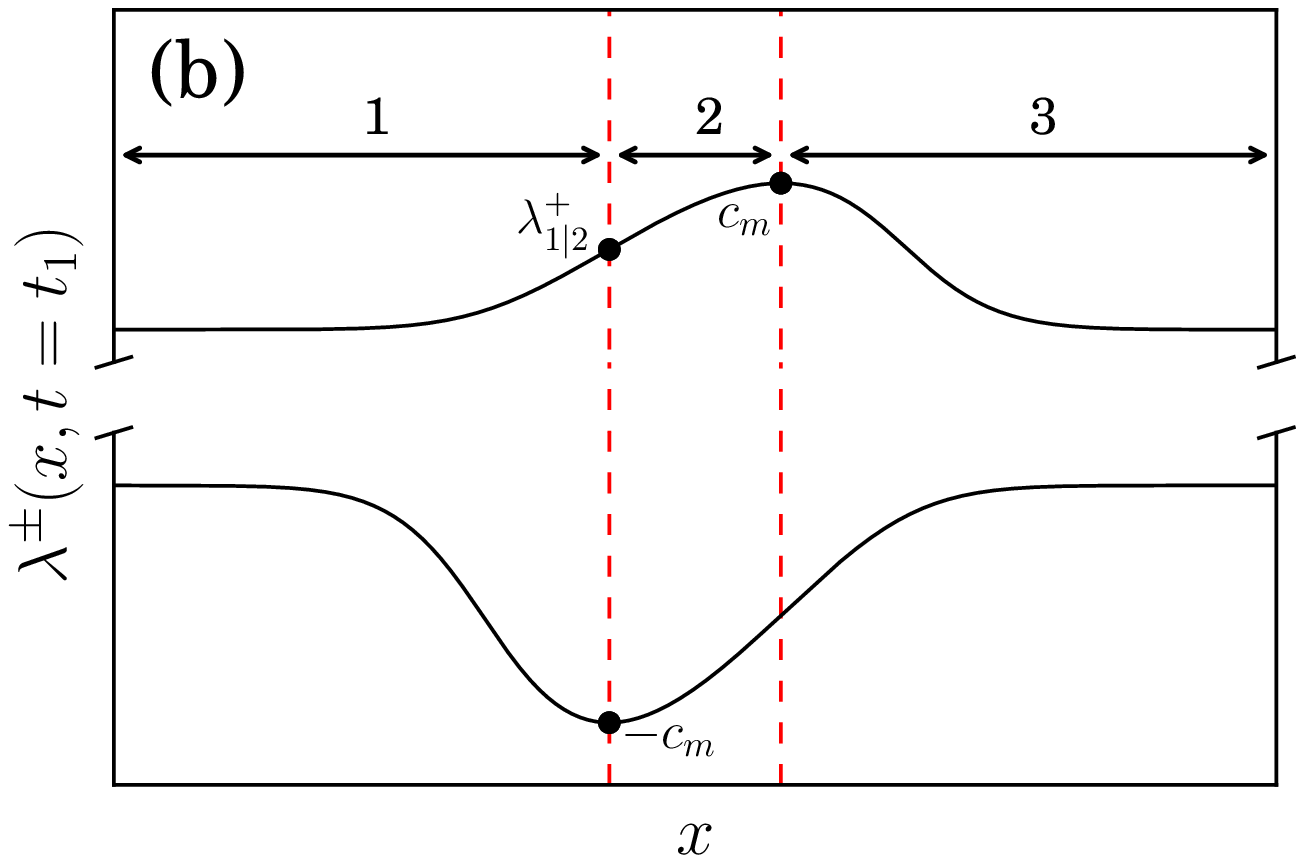}
\caption{Sketch of the distributions $\lambda^\pm(x,t)$ at time $t=0$
  (left panel) and at finite time $t>0$ (right panel).  In each panel
  the upper solid curve represent $\lambda^+$ (always larger than
  $c_0$), and the lower one $\lambda^-$ (always lower than $-c_0$).
  Panel (a) corresponds to the initial distribution, in which part B
  corresponds to region 1 and part A to region 3 (see the text).  For
  $t>0$, $\lambda^+$ ($\lambda^-$) moves to the right (to the left)
  and part B of $\lambda^+$ starts to overlap with part A of
  $\lambda^-$. This leads to the configuration represented in
  panel (b) where a new region (labeled region 2) has appeared. For
  later convenience, the value $\lambda^+_{1|2}(t_1)$ is added in this
  panel. It corresponds to the value of $\lambda^+$ at
  the boundary between regions 1 and 2 (see the discussion in
  Sec. \ref{isoard_Results}).}
\label{isoard_fig1}
\end{figure}

We first schematically depict in Fig.~\ref{isoard_fig1}(a) the initial
spatial distributions $\lambda^\pm(x,0)$ of the Riemann invariants
(left panel). The initial condition \eqref{isoard_rho_init} yields
$\lambda^\pm(x,0)=\pm \sqrt{\rho(x,0)}$. A later stage of evolution is
shown in the right panel of Fig.~\ref{isoard_fig1}. We introduce
notations for some remarkable values of the Riemann invariants:
$\lambda^{\pm}(-\infty,t) = \lambda^\pm(\infty,t)= \pm\sqrt{\rho_0}
\equiv \pm c_0$ and $\lambda^\pm(0,0)= \pm \sqrt{\rho_m}\equiv\pm
c_m$. We also define as part A (part B) the branch of the distribution
of the $\lambda^\pm$'s which is at the right (at the left) of the
extremum $\pm c_m$.  These notations are summarized in
Fig. \ref{isoard_fig1}(a).

At a given finite time, the $x$ axis can be considered as divided in
three domains, each requiring a specific treatment. Each domain is
characterized by the behavior of the Riemann invariants. In domain 3
(domain 1 respectively), $\lambda^+$ is decreasing (increasing) while
$\lambda^-$ is increasing (decreasing); in domain 2 both are
increasing, see Fig.~\ref{isoard_fig1}(b). The theoretical description
of this nonlinear wave is challenging because in each regions {\it
  both} Riemann invariants ($\lambda^+$ and $\lambda^-$) depend on
position (i.e., there is no simple wave region).

\begin{figure}[h!]
\centering
\includegraphics[scale=0.6]{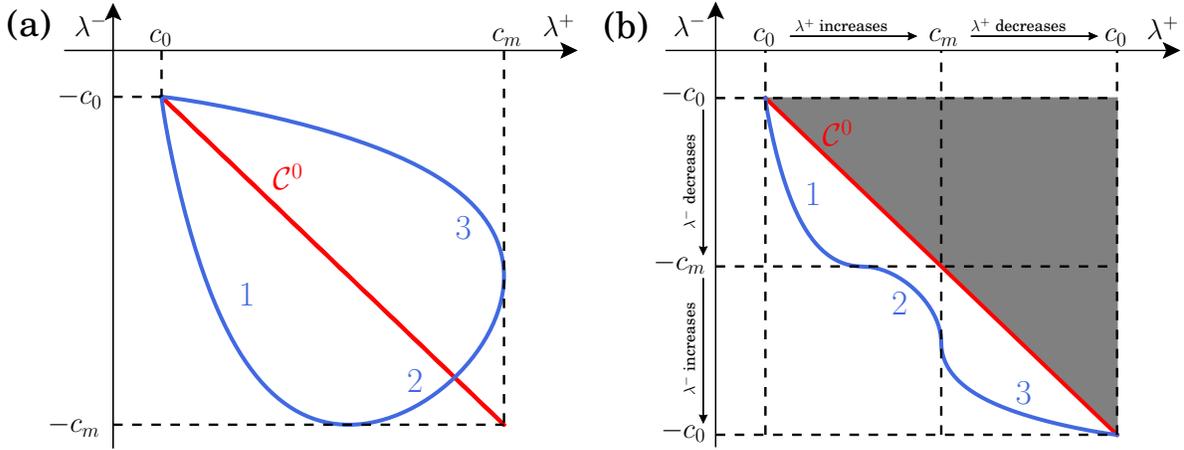}
\caption{(a) Behavior of the Riemann invariants in the characteristic
  plane at a given time $t$ (blue curve). The red curve
  $\mathcal{C}^{\, 0}$ corresponds to the initial condition
  [$\lambda^-(x,0) = - \lambda^+(x,0)$]. (b) The same curves in the
  four-sheeted unfolded surface.  In our problem, the whole gray
  shaded domain above ${\cal C}^0$ is unreachable.}
\label{isoard_fig2}
\end{figure}

The values of the Riemann invariants corresponding to
Fig.~\ref{isoard_fig1}(b) are represented in the characteristic plane
in Fig.~\ref{isoard_fig2}(a). The red curve $\mathcal{C}^{\, 0}$ in
Figs. \ref{isoard_fig2}(a) and (b) corresponds to the initial
conditions depicted in Fig. \ref{isoard_fig1}(a). Since
$\lambda^+(x,0) = - \lambda^-(x,0)$, the curve $\mathcal{C}^{\, 0}$
lies along the anti-diagonal in the characteristic plane.  The (blue)
curvy lines correspond to regions where both Riemann invariants depend
on position: the domains 1, 2 and 3. In each of these three domains
the solution $\chi$ of the Euler-Poisson equation has a different
expression. In order to describe these three branches, following
Ludford~\cite{isoard_Lud52}, we introduce several sheets in the
characteristic plane by unfolding the domain
$[c_0,c_m]\times [-c_m,-c_0]$ into a four times larger region as
illustrated in Fig.~\ref{isoard_fig2}(b). We remark here that the
whole region above $\mathcal{C}^{\, 0}$ --- shaded in
Fig.~\ref{isoard_fig2}(b) --- is unreachable for the initial
distribution we consider: for instance, the upper shaded triangle in
region 1 would correspond to a configuration in which
$\lambda^+_{{\rm region} 1}(x,t) > |\lambda^-_{{\rm region} 1}(x,t)|$,
which does not occur in our case, see Fig.~\ref{isoard_fig1}(b).  The
potential $\chi(\lambda^-,\lambda^+)$ can now take a different form in
each of the regions labeled as 1, 2 and 3 in Fig.~\ref{isoard_fig2}(b)
and still be considered as single-valued. In each of the three
domains, we use Riemann-Ludford method to solve
Eq.~\eqref{isoard_eq:EP}. This yields, to a very good approximation (a
thorough analysis can be found in Ref.~\cite{isoard_IKP})
\begin{equation}
\chi^{(n)}(\lambda^-,\lambda^+) =
\frac{\sqrt{2}}{\sqrt{\lambda^+-\lambda^-}}
\int_{-\lambda^-}^{\lambda^+} \!\!\!\! \sqrt{r}\,
\,w^{{\rm\scriptscriptstyle A}/{\rm\scriptscriptstyle B}}(r)\,dr \; ,
\label{isoard_chi1}
\end{equation}
for regions $n=1$ and $3$. In the above formula, the superscript A
should be used when $n=3$, and the superscript B when $n=1$, and
$w^{\rm\scriptscriptstyle A}$ ( $w^{\rm\scriptscriptstyle B}$) is the
inverse function of the initial $\lambda$ profiles in part A (part
B). For the initial condition \eqref{isoard_rho_init}
one has
\begin{equation*}
x = w^{\rm\scriptscriptstyle A/B}(\lambda)=\pm x_0
\sqrt{-\frac{1}{2}
\ln
\frac{\displaystyle \lambda^2-\rho_{0}}{\displaystyle \rho_1}}
\quad \text{if} \;\; x\gtrless 0 \; .
\end{equation*}
In region 2, the formulae \eqref{isoard_chi1} are replaced by
\begin{equation}\label{isoard_chi2}
{\chi}^{(2)}(\lambda^-,\lambda^+)  = \frac{\sqrt{2}}{\sqrt{\lambda^+-\lambda^-}}
\left(
\int_{c_m}^{\lambda^+} \sqrt{r}\,w^{\rm\scriptscriptstyle B}(r)\,dr
+  \int_{-\lambda^-}^{c_m}
\sqrt{r}\,w^{\rm\scriptscriptstyle A}(r)\,dr \right) \; .
\end{equation}

\subsection{Results and comparison with numerical
  simulations} \label{isoard_Results}

Once $\chi^{(n)}(\lambda^-,\lambda^+)$ has been
determined in each of the three regions ($n=1$, 2 or 3), the
problem is solved. One first computes
$W^{(n)}_\pm(\lambda^-,\lambda^+)$ in each region from
Eqs. \eqref{isoard_eq:pot}, \eqref{isoard_chi1} and
\eqref{isoard_chi2}. Then, the procedure to obtain the values of
$\lambda^+$ and $\lambda^-$ as functions of $x$ and $t$ is the following:

\bigskip $\bullet$ One starts by determining the value of $\lambda^+$
for which $\lambda^-=-c_m$ at time $t$. This value of $\lambda^+$
defines the boundary between regions 1 and 2. We denoted it as
$\lambda^+_{1|2}(t)$; it is represented in
Fig.~\ref{isoard_fig1}(b). From Eqs.~\eqref{isoard_hodograph},
$\lambda^+_{1|2}(t)$ is a
solution of
\begin{equation}
\frac{W_+^{(1)}(-c_m,\lambda^+_{1|2})-W_-^{(1)}(-c_m,\lambda^+_{1|2})}
{v_+(-c_m,\lambda^+_{1|2}) - v_-(-c_m,\lambda^+_{1|2})} + t = 0\; .
\end{equation}
We then know that, in region 1 at time $t$, $\lambda^+$ takes all
possible values between $c_0$ and $\lambda^+_{1|2}(t)$.

$\bullet$ One picks a value of $\lambda^+$ in $[c_0,c_m]$. From Eqs.
\eqref{isoard_hodograph}, $\lambda^-$ is then solution of
\begin{equation}
\frac{W_+^{(n)}(\lambda^-,\lambda^+)-W_-^{(n)}(\lambda^-,\lambda^+)}
{v_+(\lambda^-,\lambda^+) - v_-(\lambda^-,\lambda^+)} + t = 0\; ,
\label{isoard_eq_dispn}
\end{equation}
with $n=1$ if $\lambda^+ \in [ c_0, \lambda^+_{1|2}(t)]$ and $n=2$ if
$\lambda^+ \in [\lambda^+_{1|2}(t),c_m]$.  This determines the value
of the Riemann invariants in regions 1 and 2. In region 3 one uses the
symmetry of the problem and writes
$\lambda^\pm(x,t) = -\lambda^\mp(-x,t)$, see Fig.~\ref{isoard_fig1}(b).

$\bullet$ At this point, for each value of $t$ and
$\lambda^+$ we know the value of the other Riemann invariant
$\lambda^-$. The position $x$ is then simply obtained by either one of
Eqs.~\eqref{isoard_hodograph}. So, for given $t$ and $\lambda^+$ in
region $n$, one has determined the values of $\lambda^-$ and $x$. In
practice, this makes it possible to associate a couple
$(\lambda^-,\lambda^+)$ to each $(x,t)$.
The density and velocity profiles are then obtained through
Eqs.~\eqref{isoard_3-3b}.

\begin{SCfigure}[40][h]
\centering
\includegraphics[scale=0.7]{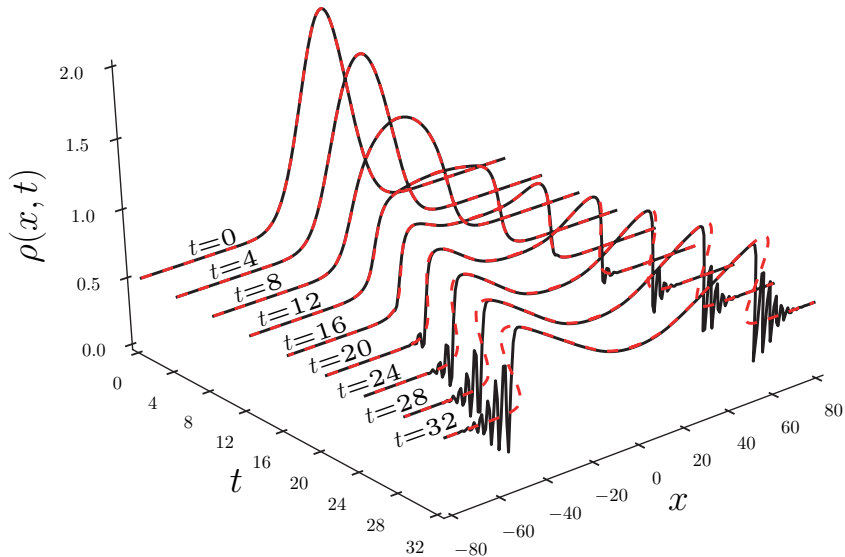}
\caption{Density profile $\rho(x,t)$ corresponding to the initial
  conditions \eqref{isoard_rho_init} with $\rho_0=0.5$, $\rho_1=1.5$
  and $x_0=20$. The red dashed line corresponds to the exact solution
  of the dispersionless system \eqref{isoard_3-3c1} (see the text),
  while the black curve displays the density obtained from the
  numerical solution of Eq.~\eqref{isoard_eq:nls}.}
\label{isoard_fig3}
\end{SCfigure}

\bigskip The results of the above approach are compared in
Fig.~\ref{isoard_fig3} with the numerical solution of
Eq.~\eqref{isoard_eq:nls}, taking the initial condition
given by Eq.~\eqref{isoard_rho_init} with $\rho_0=0.5$,
$\rho_1=1.5$ and $x_0=20$. One reaches an excellent agreement for the
density profile and also for the velocity profile (not shown in the
figure) up to $t\simeq 20$.
As time increases, the profile steepens and oscillations become
visible at both ends of the pulse at $t\gtrsim 16$.  There exists a
certain time, the wave breaking time $t_{\rm\scriptscriptstyle WB}$,
at which nonlinear nondispersive spreading leads to a gradient
catastrophe; our approximation subsequently predicts a nonphysical
multivalued profile, as can be seen in Fig.~\ref{isoard_fig3} (for
$t>20$). The time $t_{\rm\scriptscriptstyle WB}$ can be computed by
noticing that the wave breaking occurs for the value
$\lambda^+_{\rm\scriptscriptstyle
  WB}=(\rho_0+\rho_1/\sqrt{e}\,)^{1/2}$ which is associated in the
initial profile with the largest gradient in $\partial_x\rho$.  At the
wave-breaking time the profile of $\lambda^+$ in region 3 has a
vertical tangent line: $\partial x/\partial\lambda^{+}= 0$.  For
simplicity we also assume that the wave breaking occurs in a region
where one can safely approximate $\lambda^-=-c_0$. Differentiation of
\eqref{isoard_eq:hodo1} then yields
\begin{equation}\label{isoard_tWB-est0}
t_{\rm\scriptscriptstyle WB}=
\frac23
\left|\frac{dW^{(3)}_+(-c_0,\lambda^+)}{d\lambda^+}
\right|_{\lambda^+_{\rm\scriptscriptstyle WB}}\!\!\!=\left|
\frac{\int_{c_0}^{\lambda^+}
\!\!\!\sqrt{r}w^{\rm\scriptscriptstyle A}(r) dr}{\sqrt{2}(\lambda^++c_0)^{5/2}} +
\frac{\sqrt{2}(c_0-\lambda^+)w^{\rm\scriptscriptstyle A}(\lambda^+)}
{3\sqrt{\lambda^+}(\lambda^++c_0)^{3/2}}
+\frac23\sqrt{\frac{2\lambda^+}{\lambda^++c_0}}
\frac{dw^{\rm\scriptscriptstyle A}}{d\lambda^+}
\right|_{\lambda^+_{\rm\scriptscriptstyle WB}} .
\end{equation}
The numerical value of $t_{\rm\scriptscriptstyle WB}$ is found to be
$\simeq 19.15$ for our choice of initial condition, in good
agreement with numerical simulations. Note also that for a small bump
($\rho_1\ll \rho_0$) the wave breaking time becomes very large.  From
\eqref{isoard_tWB-est0}, and for an initial profile of type
\eqref{isoard_rho_init}, one gets at leading order in $\rho_1/\rho_0$:
\begin{equation}
t_{\rm\scriptscriptstyle WB} \simeq \frac{2\sqrt{e}}{3}
\frac{x_0}{c_0} \left(\frac{\rho_0}{\rho_1}\right) \; .
\end{equation}
This means that the breaking time is much greater than the time $\sim x_0/c_0$
of propagation of sound along the pulse profile.
In our optical system the wave breaking is regularized by the
formation of a dispersive shock wave which is a region with large
oscillations of intensity and phase, whose extend increases with time,
as can be seen in Fig.~\ref{isoard_fig3}.  Its description requires a
nonlinear treatment able to account for dispersive effects and this
goes beyond te scope of the present letter (see, e.g.,
Ref.~\cite{isoard_IKP}).

\section{Conclusion} \label{isoard_conclusion}

In this work we demonstrate that a nondispersive hydrodynamic approach
to the spreading and splitting of an optical pulse compares extremely
well with the results of numerical simulations up to the wave breaking
time. At larger time, one observes the formation of an optical
dispersive shock wave, which can be studied within Whitham modulation
theory. In the case of the initial distribution given by
Eq.~\eqref{isoard_rho_init}, the shock should be described by four
varying Riemann invariants and this requires a thorough
investigation. Work in this direction is in progress.

\end{document}